\documentclass[
    ,final  ]
  {aipproc}

\layoutstyle{6x9}
\usepackage{graphicx}

\begin{document}

\title{The ANTARES Neutrino Telescope: status and first results.}

\classification{95.55.Vj, 95.85.Ry}
\keywords      {ANTARES, neutrino telescope, dark matter, point source}

\author{Anthony M Brown, on behalf of the ANTARES Collaboration}{
  altaddress={brown@cppm.in2p3.fr}, address={CPPM-Centre de Physique des Particules de Marseille, 164, Av. de Luminy, Case 902, Marseille, France.},altaddress={http://antares.in2p3.fr}  
}

\begin{abstract}
Completed in May 2008, the ANTARES neutrino telescope is located in the Mediterranean Sea, 40 km off the coast of Toulon, at a depth of about 2500 m. Consisting of 12 detector lines housing nearly 900 optical modules, the ANTARES telescope is currently the largest neutrino detector in the northern hemisphere. Utilising the Mediterranean Sea as a detecting medium, the detection principle of ANTARES relies on the observation of Cherenkov photons emitted by charged relativistic leptons, produced through neutrino interactions with the surrounding water and seabed, using a 3 dimensional lattice of photomultiplier tubes. In this paper we review the current status of the ANTARES experiment, highlighting some of the results from it's first year of full operation.
\end{abstract}

\maketitle

%%%%%%%%%%%%%%%%%%%%%%%%%%%%%%%%%%%%%%%%%%%%
%% MAINBODY
%%%%%%%%%%%%%%%%%%%%%%%%%%%%%%%%%%%%%%%%%%%%

\section{Introduction}

Neutrinos afford us a unique view of the Universe. Due to their small interaction cross-section and neutral charge, neutrinos are able to escape, undeflected, from some of the most dense regions of the Universe. As such, neutrinos offer us a unique probe to the most extreme processes occuring in cosmic sources. 

The production of high energy neutrinos can occur via the decay of charged mesons, with the charged mesons being produced through the interaction of accelerated hadrons with ambient matter or radiation fields. In an astrophysical scenario, the production of high energy neutrinos is believed to occur in numerous sources where hadronic, or Cosmic Ray, acceleration is believed to take place, such as active galactic nuclei, supernova remnants and microquasars \cite{source1}\cite{source2}\cite{source3}. As such, neutrino astronomy plays an important role, in the context of a multi-messenger approach, in conclusively proving the origin of Cosmic Ray particles. 

Searching for neutrino signatures from Dark Matter annihilation is another key goal of neutrino telescopes. Through elastic scattering, weakly interacting massive particles (WIMPs), relic from the Big Bang, are believed to accumulate at the centre of massive celestial objects such as the Sun. This over density of WIMPs results in an increase in the self-annihilation rate of these particles into ordinary matter, producing a measureable high energy neutrino flux in the process.

With a view of pursuing these two main scientific goals, point source and indirect dark matter searches, the ANTARES collaboration has built the largest neutrino telescope in the northern hemisphere, off the coast of southern France. After reviewing the detection technique and performance of the ANTARES neutrino telescope, we will discuss these two scientific goals in light of ANTARES' first year of complete 12 line operation.

\section{The ANTARES telescope}

The ANTARES neutrino telescope consists of 12 detector lines, each approximately 480 meters in length, covering a footprint of approximately $0.1$ km$^2$ \cite{carr}. The physical dimensions of ANTARES have been optimised for the detection of up-ward going relativistic muons produced through charged-current interactions between muon neutrinos and the water/bedrock surrounding the detector. As such, each detector line supports 75 10-inch photomultiplier tubes housed in pressure resistent glass spheres (OMs), distributed over 25 storeys \cite{om}. Each storey contains a triple of OMs looking downwards at an angle of 45$^\circ$ from the horizontal. The vertical spacing between each storey is 14.5 meters, with the first storey located 100 meters above the seabed. 

The detection technique of the ANTARES telescope relies on the observations of Cherenkov photons emitted by relativistic charged leptons produced through neutrino interactions with the surrounding water and seabed, using a 3 dimensional lattice of photon detectors. The lepton's track through the detector is reconstructed using the timing and position of the Cherenkov photon hits on the 3-D lattice. The ANTARES telescope has an energy threshold of ~20 GeV for reconstructed neutrino events. However, the telescope's performance dramatically improves with increasing neutrino energy such that at a neutrino energy of 10 TeV, the ANTARES telescope has an effective area of ~0.5 m$^2$ and an angular resolution better than 0.3$^\circ$. The performance of these important detector characteristics, as a function of neutrino energy, can be seen in Figure 1.

\begin{figure}
  \includegraphics[height=.2\textheight]{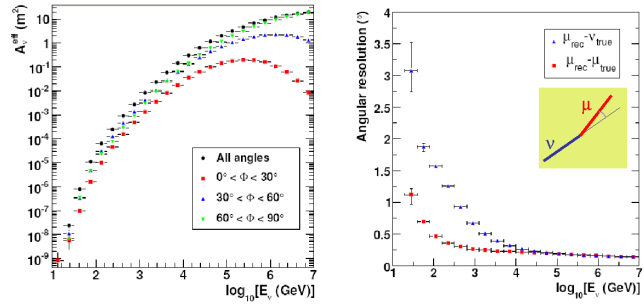}
  \caption{\emph{Left:}The effective area of the ANTARES telescope, as a function of neutrino energy. It should be noted that above approximately 1 PeV, the effective area starts to decrease due to the Earth becoming opaque to neutrinos above this energy. \emph{Right:} The angular resolution of the ANTARES telescope as a function  of energy.}
\end{figure}

The ANTARES telescope started taking data with the connection of the first line in March 2006 \cite{firstline}. By January 2007, a total of 5 detector lines had been connected and the search for point sources and dark matter neutrino signals commenced. With the detector doubling in size at the start of December 2007, ANTARES was completed at the end of May 2008, with the connection of the final 2 detector lines.

\section{Atmospheric Muons}

Even at a depth of 2500 meters under water, atmospheric muons, produced by cosmic ray interactions in the atmosphere, are the dominant signal in the ANTARES detector, with a factor of $10^5$ between triggers from atmospheric muons and atmospheric neutrinos. Nonetheless, atmospheric muons afford us an excellent opportunity to check the systematic uncertainties of the detector. 

Using all 5 line data for 2007, the azimuth and zenith distribution of the atmospheric muon triggers were compared to the Monte Carlo (MC) simulations of the detectors response. These MC simulations take into consideration both detector properties, such as angular acceptance of the OMs, and the uncertainties associated with the atmospheric muon production and propagation. As can be seen in Figure 2, there is a good agreement between the data and our MC predictions. This agreement, within the uncertainties, indicating that we understand the properties and the systematic uncertainties of the detector. It should be noted that the zenith distribution of atmospheric muons is dominated by down-going events, and as such, to remove a major source of background, ANTARES only considers up-going events as neutrino candidates. The uneven distribution of the azimuth distribution of atmospheric muons is due to the detector line distribution on the seabed. 

\begin{figure}
  \includegraphics[height=.2\textheight]{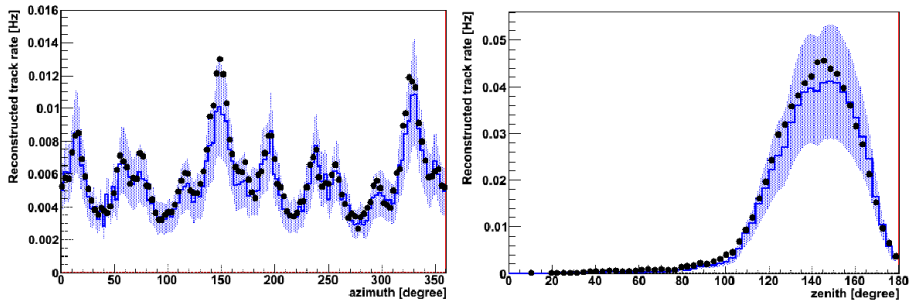}
  \caption{The azimuth (left) and zenith (right) distribution of the atmospheric muon triggers, with the black dots being the data, the solid line being the MC predictions and the band being the uncertainty associated with these MC predictions.}
\end{figure}

\section{Cosmic Neutrino Point Source Search}

To minimise the effect of atmospheric muons, we only consider up-going events to be neutrino candidates. As such, ANTARES observes the southern sky, and thus, the majority of the galactic plane. From a `point source' point of view this is important since work by the H.E.S.S. collaboration have shown the inner galaxy to be populated with many objects capable of accelerating particles to the TeV energy range needed for TeV neutrino emission \cite{hess}. 

The 5 line point source search was performed using two independent and complementary search algorithms: a powerful, unbinned, iterative Expectation-Maximazation method \cite{em1} \cite{em2} and a binned, cone search algorith using Poisson statistics \cite{p1}. Both of these methods were applied using two approaches. Firstly we searched for a neutrino signal from a list of 25 possible source candidates. This list of potential neutrino sources included supernova remnants, microquasars, active galactic nuclei and our Galactic Center, and took into consideration the very high energy $\gamma-$ray properties of the source as well as the total visibility of the source to the ANTARES detector. Considering a list of potential candidates allowed us to improve the sensitivity of our search since we restrict the number of locations in the sky that is being investigated. The second approach involved a blind search across the whole sky. While this second approach is less sensitive, it is necessary to ensure that clusters of neutrino candidates from other sources not included in the list, are not overlooked. Nonetheless, applying both of these methods, to both search approaches, no neutrino point source was found in the 5 line 2007 data, at 90\% confidence level. 

The upper limits derived from this 5 line data set, assuming an $E_{\nu}^{-2}$ spectrum, for the list of 25 candidates, are shown in Figure 3. It should be noted that, while using a data set of only 140 days and using a partially completed detector, ANTARES was able to set upper limits that are comparable, if not better than those set by other southern looking neutrino experiments with multi-year data sets.

\begin{figure}
  \includegraphics[height=.25\textheight]{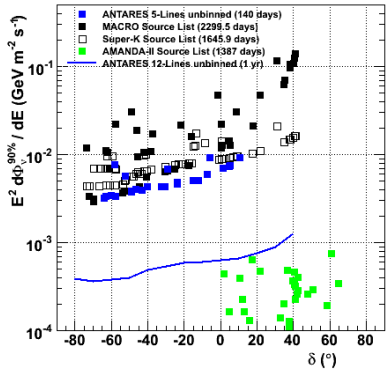}
  \caption{Neutrino flux upper limits (solid squares) set by the 5 line configuration of ANTARES, for a list of 25 candidate source, compared with results from other experiments (IceCube\cite{ic1}, AMANDA\cite{ic2}, SuperKamiokande \cite{sk1} and MACRO \cite{mac1}), at 90\% confidence level, assuming an $E_{\nu}^{-2}$ spectrum. The predicted sensitivity of one year ANTARES data set is also shown.}
\end{figure}

\section{Indirect Dark Matter Search}

Within the supersymmetric minimal supergravity (SUSY mSUGRA) framework, the neutralino, $\chi$, can be the lightest stable SUSY particle. With a mass in the range of GeV to TeV, the neutralino interacts through gravity and the weak force alone, and thus it fulfils all the requirements to be a cold dark matter candidate. In the context of neutrino telescopes, searching for WIMP dark matter candidates involves looking for neutrino fluxes from the core of massive celestial objects such as the Sun, our Galaxy or indeed from the centre of the Earth, where the WIMPS are believed to accumulate. 

\subsection{ANTARES sensitivity to Neutralino Annihilation}

The expected neutrino flux from neutralino annihilations in the Sun, in the mSUGRA framework, depends upon four parameters: $A_0$, $m_0$, $m_{1/2}$ and $tan(\beta)$, where $A_0$ is the universal trilinear coupling, $m_0$ is the universal scalar mass, $m_{1/2}$ is the universal gaugino mass and $tan(\beta)$ is the ratio of the vacuum expectation values of the two Higgs fields. Using a modified version of the DarkSUSY code \cite{darks1}, the expected neutrino flux was calculated for approximately 4 million different combinations of these four parameters, with each different parameter set being chosen using a random walk method through the parameter space. It should be noted that this flux calculation also takes into consideration neutrino oscillations and absorption within the Sun. 

Combining these calculated neutrino fluxes with the energy dependent effective area of ANTARES, we can estimate a detection rate for dark matter induced neutrino emission. This estimation is essentially a indication of ANTARES' ability to probe neutralino properties, such as relic density or mass, and can be seen in Figure 4, along with predictions for the future kilometer cube size expansion (KM3NET \cite{km3v2}).

\begin{figure}
  \includegraphics[height=.25\textheight]{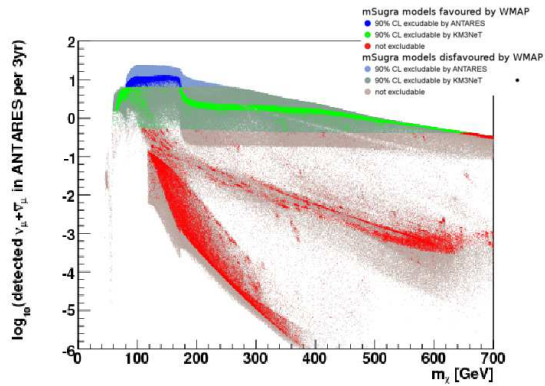}
  \caption{Sensitivity of the ANTARES and KM3NET to neutrinos produced through neutralino annihilation in the Sun, in the mSUGRA framework. The neutrino flux is integrated above 10 GeV, as a function of neutralino mass. The blue and green data indicate areas of the model parameter space accessible by ANTARES and KM3NET respectively, while the red points are inaccessible to both experiments. The bright data points indicate models that are predicted, within 2$\sigma$ of the WMAP relic density, the shaded points are those outside of this.}
\end{figure}

\subsection{ANTARES neutrino flux limit from dark matter annihilation in the Sun.}

The 5 line data from 2007 was also used to search for neutrino signals from neutralino annihilation in the Sun. Only up-going events when the Sun was below the horizon were considered, resulting in an effective lifetime of 68.4 days. Considering two extreme cases for the neutralino annihilation channel, namely hard (W-boson) and soft (b quark), the size of the search cone around the Sun was optimised before being applied to the data. Using this search cone, limits on the neutrino flux from the Sun, above 10 GeV, were calculated for both the hard and soft channels, following the unified approach method of Feldman and Cousins \cite{fc1}. These limits can be see in Figure 5.  

\begin{figure}
  \includegraphics[height=.25\textheight]{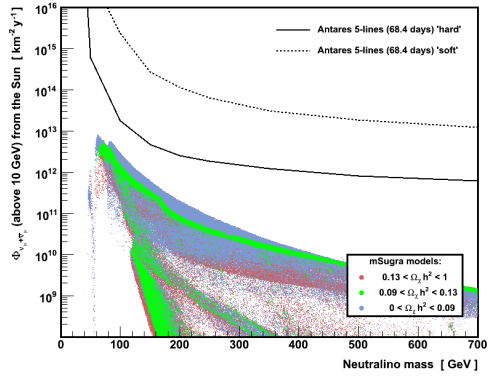}
  \caption{The upper limit on the neutrino flux from neutralino annihilation in the Sun, assuming two extreme cases for the annihilation channel. Set for neutrinos above 10 GeV in energy, these limits were calculated using all 5 line data from 2007.}
\end{figure}

To compare ANTARES' neutrino flux limits from neutralino annihilation, to those set by other experiments, the neutrino flux limit derived in Figure 5 was converted to the corresponding muon flux, integrated above an energy of 1 GeV. These comparisons can be seen in Figure 6. While the limits set by ANTARES in Figure 6 are not as constraining as the other experiments, it should be noted that ANTARES' flux limits are based on only approximately 70 days of data from the less than half completed detector.

\begin{figure}
  \includegraphics[height=.3\textheight]{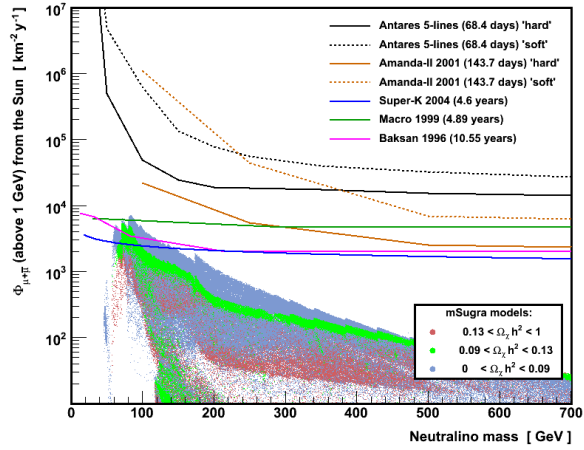}
  \caption{The neutrino induced muon flux from the upper limits set in Figure 5, intergrated above a muon energy of 1 GeV. As before, the green data points are from a relic density that is within 2 $\sigma$ of the WMAP region, with the red and blue show relic densities above and below this relic density respectively.}
\end{figure}

\section{Conclusion}

The ANTARES neutrino telescope was completed in May 2008 with the connection of the last two detector lines. The flux and distribution of atmospheric muons detected by ANTARES have shown that we understand the systematic uncertainties of the now completed telescope. Throughout its construction, ANTARES has been taking data. Data taken during its 5 line configuration stage (March to December 2007) has been used to set competitive flux limits on a list of possible neutrino point sources. Assuming an $E_{\nu}^{-2}$ neutrino spectrum, the upper limits set by ANTARES' 5 line data are better then those set by the multi-year data sets of Super-Kamiokande and MACRO experiments. This improvement being a testiment to ANTARES' unprecedented sensitivity to neutrinos in the 10-100 TeV energy range. 

The 5 line data was also used to set limits on the neutrino flux from possible neutralino annihilation in the center of the Sun, assuming both hard and soft annihilation channels. While these upper limits are non-constraining, a similiar study using a data set from the now completed ANTARES detector will allow us to place better limits in the not-to-distant future. 

\begin{theacknowledgments}
Anthony Brown acknowledges the financial support of both the French funding agency `Centre National de la Recherche Scientifique' (CNRS) and the European KM3NET design study grant.
\end{theacknowledgments}

\bibliographystyle{aipprocl}

\end{document}